# Modeling clustered non-stationary Poisson processes for stochastic simulation inputs

Issac Shams, Saeede Ajorlou, Kai Yang

*Department of Industrial and Systems Engineering, Wayne State University, Detroit, MI*

**Abstract**

A validated simulation model primarily requires performing an appropriate input analysis mainly by determining the behavior of real-world processes using probability distributions. In many practical cases, probability distributions of the random inputs vary over time in such a way that the functional forms of the distributions and/or their parameters depend on time. This paper answers the question whether a sequence of observations from a process follow the same statistical distribution, and if not, where the exact change points are, so that observations within two consecutive change points follow the same distribution. We propose two different methods based on likelihood ratio test and cluster analysis to detect multiple change points when observations follow non-stationary Poisson process with diverse occurrence rates over time. Results from a comprehensive Monte Carlo study indicate satisfactory performance for the proposed methods. A well-known example is also considered to show the application of our findings in real world cases.

## 1. Introduction

Simulation input data analysis is often considered as a vital step in most stochastic simulation experiments enabling analysts to drive reasonable models from real-world systems. The ultimate goal of input data analysis is selection of valid input models, mainly the statistical distributions that appropriately represent the behavior of the system under consideration. In the simplest case, an input parameter is assumed to be independent and identical random variable (i.i.d), which follows a known and standard distribution. However, in many real-world cases, there is no guarantee that such assumptions are always met. Vincent (1998) noted that there are many alternatives to these assumptions that can be addressed in different applications. For instance, the input data may be correlated which means the consecutive observations depend on each other. In fact, the observation *t* can be modeled with a linear or nonlinear combination of the last observations *t*-1, *t*-2,… together with an independent white noise.

In addition, a data set can be inherently independent but, as Vincent (1998) mentioned, it is not stable over time. In other words, the functional forms of the input distributions and/or their parameters vary over time. A well-known case occurs when a random input variable follow a non-stationary (time-dependent) process in which its distribution changes in such a way that the basic functional form is stable, but it has parameter(s) that depend on time. Kuhl et al. (2009) refers to this problem as *parameter uncertainty*. For example, in many applications, customer arrival rates in rush hours are larger than the other periods. There are many special cases for non-stationary processes including the following ones:

- A data set from a random input variable may be *clustered* such that the distributions of the parameters vary over time but observations within a cluster follow the same distribution. For example, the arrival rate of customers at a fast food may change over a day. However, the arrival rate of customers is same during morning, and similarly the arrival rate of customers is same in afternoon.
- A data set from a random input variable may be *mixture* in which the consecutive observations may come from a distribution with different parameter(s) values. This is mainly due to the fact that the different entities in the system under study have various attributes affecting the input variables in several ways. For instance, assume that different customers in a bank queue need various services (attributes), i.e. cashing or depositing checks, opening a new account, or even paying loans. The process time for each service may be diverse and follow distributions with diverse parameters. Since people in a bank queue need different and random services, the bank teller service time is a mix of several distributions and the data set collected on such an input would follow a mixture distribution.

A general model frequently used to represent non-stationary processes, specially arrival times, is Nonhomogeneous Poisson Processes (NHPP) which has successfully been performed to model complex time-dependent arrival processes in many simulation studies (Kuhl et al., 2008). Some examples in this category include (a) Repairable systems such as fan motors and turbines (Kuhl et al. (2008)), (b) Calls for online analysis at a hospital in Houston (Kao and Chang (1998)), and (c) Respiratory cancer deaths in Scotland (Lawson (1970)). In a NHPP, it is assumed that the arrival rate λ depends on time. Hence, $N(t)$, the number of arrivals during the time interval (0,$t$], varies over time and time-dependent arrival rate denoted by λ($t$) is a nonnegative, integrable function satisfying the usual Poisson assumptions. In case that the arrival rate exhibits a strong dependence or a specific and complex pattern i.e. cyclic, nested cyclic, or trend patterns, researchers often estimate the mean-value function E[N($t$)] over time using different parametric and nonparametric methods. Lee et al. (1991), Kuhl et al. (1997), and Kuhl and Wilson (2000) addressed the estimation of mean-value function using parametric methods. Lewis and Shedler (1976), Pritsker et al. (1995), Leemis (1991; 2000; 2004), Kuhl and Wilson (2001), Kuhl et al. (2006), Alexopoulos et al. (2008), and Kuhl et al. (2008) proposed different nonparametric or semiparametric methods to estimate the mean-value of a NHPP as a function of $t$. More recently, Kuhl and Wilson (2009) did a comprehensive review on various methods for modeling and simulating NHPPs in simulation applications.

As mentioned, estimating the mean-value is an appropriate method for modeling NHPPs, particularly when the arrival rate depends strongly on time represented by a specific and/or complex pattern. These methods are basically curve fitting practices and attempt to fit an estimator function on the values of $\mu(t)$ in a given time interval in a given time interval $[0, S]$, in various situations such as sinusoidal oscillations, cyclic pattern, long-term trend, or nested periodic phenomena.

Alternatively, another approach used to model such a time-dependent process is to divide the time interval (0,$t$] to a finite number of disjoint subintervals and estimate $\lambda_i$, not the mean value $\mu(t)$, for the $i^{th}$ subinterval. This approach is specifically useful when observations in a data set are clustered. Law (2007) discussed that this method is a heuristic but practical approach; however, there is still an important unanswered question on how one decides on the length of subintervals. Although a graphical method for approximately determination of the subinterval bounds is proposed, but the performance of the suggested method is not investigated through the use of statistical measures.

In this paper, we propose two different approaches capable of detecting the cluster patterns in a given data set and estimating the change points (the subintervals of length) for each cluster. The first method is based on nonparametric hierarchical cluster analysis using a dissimilarity measure and a specified algorithm. The second method is based on a likelihood ratio test which helps to quickly detect sustained changes in a historical data set and identify the exact location of change points. The cluster analysis can be generalized to detect the presence of one or more changes in the functional form of the data where the first and higher moments of observations vary over time. Therefore, it can be used for any simulation input data rather than arrival random variables. We compare the two proposed methods using a comprehensive simulation study and different performance measures, i.e. change point locations and dispersions and number of detected changes.

The rest of this paper is organized as follows. In Section 2, the problem is discussed in details and a basis for applying change point detection techniques in simulation input analysis is provided. Both cluster analysis method and likelihood ratio test are presented in Section 3. In Section 4, the performance of the proposed methods based on accuracy and precision measures are evaluated. Section 5 discusses a well known example with the aim of demonstrating the applicability of the proposed approaches. Our concluding remarks are presented in the final section.

## 2. Multiple Change Point Detection Methods

Change point detection techniques have successfully been used in numerous statistical methods including regression analysis, statistical process control, signal processing, and pattern recognition. These techniques typically deal with identifying the time that a change occurs in the observations of a data set collected (or being accumulated) over time. If a data set has one or more change points, at least a part of observations has different moment(s) and/or different distribution(s) from the rest of observations. Knowing the exact time of a change point helps researchers and practitioners to have more accurate and realistic insight about the system they study. Change point problem has extensively been studied by many researchers. For instance, Chen and Gupta (2000) considered the single and multiple change point problems for most discrete and continuous probability distributions using likelihood ratio test, Bayesian, and information approach methods. Besides the parametric method, various nonparametric methods such as Hidden Markov Model (HMM) and Neural Networks (NN), are also proposed by many statisticians (For a detailed literature review see Brodsky and Darkhovsky (1993)). In this section, we present the general problem of change point detection for simulation input variables when observations are clustered.

A random input variable can be considered as a stochastic process, so data set $S$ can be defined as a collection of random variables $S=\{X_t; 1,2,\ldots,m\}$ over time. Note that the index $t$ represents the order of variables (discrete event variables) and $S$ is a finite set. In order to fit a specific probability distribution to data set $S$, we must check whether the observations are independent and identical. The second assumption indicates that $x_t$'s should follow the same probability distribution $F_t(.) = F(.); t = 1,2,...,m.$ Now assume that $x_t$'s are not identical but they are clustered in $R+1$ different populations. In this situation, data set $S$ can be divided into $R+1$ disjoint subgroups, $S_1$, $S_2\ldots S_{R+1}$ whose union is $S$. Besides, it is assumed that observations within $S_i$ are independent and identically distributed; however, observations from two successive sets are unlike. Let $\tau_j$ be the $j^{th}$ change point in data set $S$ where $\tau_j$; $j=1,2,\ldots,R$ is subjected to $0 < \tau_1 < \ldots < \tau_R < m$. In this paper, it is assumed that $\tau_j$ is the location of the first element of set $S_j$ which means:

$$X_t \sim F_j(.), \quad \tau_{j-1} < t \leq \tau_j; \quad j = 1,...,R. \tag{1}$$

It is of interest to estimate locations of change points $\tau_j$'s in a statistical array of data obtained (or being augmented) as a result of some industrial, simulation, or other types of experiments. Several approaches have been proposed for the case when there are multiple changes in the location parameters of random variables. Many authors employed cluster analysis to segment data at proper locations and estimate the existing change points. Most cluster analysis methods used for change point detection are based on hierarchal methods with different distance measures and optimization methods ranging from dynamic programming to metaheuristics. McGee and Carelton (1970), Hawkins (1976), and Sullivan (2002) used cluster analysis to estimate multiple change points. Hawkins (2001) extended this view using dynamic programming to find the exact maximum likelihood statistic for the general exponential family distributions. In addition, some researchers studied nonparametric change point detection methods which do not require any distributional assumptions (See, for example, Carlstein (1988), Ferger (1995), and Guan (2004)). Using single change point detection methods is another approach that can be used for the identification of multiple change points. Vostrikova (1981) pointed out that a method for detecting and estimating a single change point may be able to apply for multiple changes by binary segmentation. Once a change detected, data are divided at the estimated single change point and the procedure is applied for the two new groups. This procedure should be continued until there is no evidence for any change in all groups. Regarding this issue, Sullivan (2002) stated that for the data sets which need not to follow any single pattern or regime, the presence of multiple change point estimators may be inaccurate.

In the next section, we present the proposed methods. The first method is an extension of the cluster analysis method proposed by Sullivan (2002) for normal random variables while the second method is a binary segmentation method which is based on the likelihood ratio test for exponential random variables.

# 3. Proposed Methods

## 3.1. *Multiple Change Point Detection Using Clustering*

Suppose there are $R$ changes in the functional form of $m$ independent observations from data set $S$, and $\tau_j$ ; $j=1,2\ldots,R$ is the $j^{th}$ change point where $\tau_0=0$ and $\tau_{R+1}=m$. Let $F_t(.)$ represent the probability distribution associated with observation $x_t$. In this case, $x_t \sim F_j(.)$ , $\tau_{j-1} < t \leq \tau_j$ ; $j=1,2\ldots,R+1$ subject to $F_j(.) \neq F_{j+1}(.)$. We want to determine whether the observations are identical, $R=0$ or not; and if the observations are not identical then where the exact change points (subintervals bounds) are. The hierarchical clustering algorithm can be appropriately applied in order to answer both questions.

The algorithm typically starts with $m$ clusters, one for each observation, and ends with a single cluster including all $m$ observations (agglomerative). In each iteration, an observation or a group of observations is merged with another observation or group. Divisive algorithm may also be considered as an alternative in which all observations first include in a cluster and finally the algorithm ends with each observation inputted in a single cluster (see Tan, Steinbach, and Kumar (2005) chapter 8 for more details). The last observation in cluster $j$, denoted by $l_j$, is considered as the change point between clusters $j$ and $j+1$. Clustering methods generally use an index of *similarity* or *proximity* between each pair of observations to identify the observations that are similar and group them into clusters. A convenient measure of similarity is the distance between each pair of observations denoted by $d_j$ for adjacent clusters $j$ and $j+1$. There have been many distance functions in the literature of cluster analysis developed for different cases. However, in this paper we define $d_j$ as

$$d_j = \frac{\left|\bar{x}_j - \bar{x}_{j+1}\right|}{\sqrt{s_j^2/m_j + s_{j+1}^2/m_{j+1}}}, \quad (2)$$

where $m_j$ and $m_{j+1}$ are the sample sizes for the $j^{th}$ and $(j+1)^{th}$ adjacent clusters, $\bar{x}_j$ and $\bar{x}_{j+1}$ are the sample means, and $s_j$ and $s_{j+1}$ are the sample standard deviations, respectively. This method is similar to *centroid* method, except the distance between two clusters is defined as the "*Manhattan*" distance between the mean vectors. At the beginning of step $k$; $k=1,2,…,m-1$, there are $m-k$ boundaries and $m-k+1$ clusters of observations (Note that for step one, the denominator of $d_j$ is set equal to 1). The boundary with smallest distance is removed, and the location of the removed boundary together with its distance is updated as follows: $l^*_{m-k} = l_{k^*}$ and $d^*_{m-k} = d_{k^*}$, where $k^* = $ arg min $[d_k]$. The algorithm continues until all boundaries are removed.

Sullivan (2002) declared if the data set are stable and there is no change point in observations, then the sequence $\{d^*_j\}$ should uniformly and slowly decrease. If there are $R$ change points in data set, then $d^*_R - d^*_{R+1}$ should be large and after $d^*_{R+1}$, all distances should decrease slowly. This provides an insight to make a decision rule for recognizing multiple change points. Note that equation (2) follows a *t*-distribution with $m_j+m_{j+1}-2$ degrees of freedom if observations are independently and identically distributed normal random variables. Under this condition, Sullivan (2002) suggested using *p*-values for the test of a difference in the mean of adjacent clusters. A large *p*-value indicates that the corresponding boundary is not really a change point. However, if observations do not follow normal distribution, even if observations come from a normal population but there is a change in their higher moments, then equation (2) does not have any known distribution and *p*-value cannot appropriately be found. Alternatively, one can use pre-specified threshold value(s) for one or more $d_j^*$'s to identify single or multiple change points. For example, one can consider $g$ threshold values to check whether there exists one or more (up to $g$) change points in a random input data set. The last distance in the sequence $\{d^*_j\}$ which is greater than its threshold value (say $d_R^*$) is considered as the last change point. In this case, the estimated locations of change points are $\{l^*_j, i=1,2,…,R\}$. In other words, $l^*_q$ is an estimate for the location of $q^{th}$ change point, if $d_R^*$ is the last distance such that $d_R^* > H_R$ and $q \leq R$; $R=1,2,…,g$, where $H_R$ is the threshold value of $d_R^*$ and $g$ is the maximum potential change points that can be recognized. In order to control the false detection probability, one can set $H_R$ to be $100(1-\alpha)$ percentile of $d_R^*$. Notice that according to Bonferroni inequality, the maximum overall false detection probability for all $g$ threshold values is $\alpha g$. It should be considered that sequence $\{d_j^*\}$ has a relatively decreasing pattern and are considered to be correlated. This is why we use Bonferroni inequality to adjust the overall false detection probability.

Consider again that equation (1) is particularly appropriate when changes in cluster are due to shift(s) in location parameters while higher moments kept unchanged over time. Therefore, this assumption causes the method to be very restricted. Figure 1 presents a run chart associated with a clustered data set where the first 100 observations are independent and identical exponential random variables with parameter $\lambda=1$ while the next 100 observations follow the same distribution with parameter $\lambda=5$. It can be easily observed that as the distribution parameter increases, not only the mean but also the standard deviation of the observations change at the same time.

[Insert Figure 1 about here]

Caption: Figure 1. Run Chart of Observations

In this situation, the method may introduce some clusters with high within variation, which contradicts main objectives of clustering techniques. That is, $d_j$ may be affected by the larger standard deviation and consequently the algorithm cannot lead to good clusters. To properly overcome this obstacle, we use a variance-stabilizing transformation which does not require any distributional assumptions. In this paper, we employ the well-known Box-Cox power transformation to stabilize the second and higher moments of $X_t$ input random variables. The Box-Cox transformation (Box and Cox (1964)) for $x$'s can be written as

$$x_t(\eta) = \begin{cases} \dfrac{x_t^\eta - 1}{\eta}, & \text{if } \eta \neq 0; \\ \ln x_t, & \text{if } \eta = 0. \end{cases} \quad (3)$$

Note that different methods in the literature have been proposed to estimate the value of η. Here, we estimate η such that the standard deviation of a standardized transformed variable is minimized and as a result the within variability of each cluster is stabilized.

### 3.2. *Multiple Change Point Detection Using Likelihood Ratio Test*

Likelihood Ratio Test (LRT) has been considered as one of the most powerful tools for detecting one or more changes in a set of observations in applied statistics literature. This procedure consists of calculation of the likelihood function for all possible partitions of the data set into typically two groups. When LRT statistic exceeds a threshold value, a change point is detected and the most likely location of the change is determined by the partition corresponding to the maximum value of the statistic. The LRT method can be applied to detect not only the presence of multiple change points in an input data set, but also their locations as well. As previously stated, once a single change point is detected, the data set is divided at the estimated change point and the LRT statistic is formed separately for the two new groups.

Although most researchers asserted that LRT method outperforms many competing approaches in identifying change points but it has a restrictive assumption, i.e. knowing the exact the probability distribution of the random inputs before estimating the change points. Since LRT method relies upon distributional assumptions, the probability distribution of the input data set should be specified prior to constructing an appropriate LRT statistic. Despite the fact that the probability distribution of some input variables like interarrival time can be guessed (Law (2007)), in most practical cases, it is actually not easy to have an accurate insight about the exact distribution of the input data. For cases with exponential family densities such as gamma and Poisson, the extension of LRT is straightforward (see Hawkins (2001) for more details). However for other situations like Johnson family of distributions and Bezier densities, one may shift gears out of LRT extension and examine non-parametric approaches instead. In this paper, we consider the case that observations simply come from an independent univariate exponential distribution with rate parameter of $\lambda$.

Again assume that there is a random data set from an input variable including $m$ independent observations $x_t$'s such that $x_t \sim F_j(.)$, $\tau_{j-1} < t \leq \tau_j$; $j = 1,...,R+1$ where $\tau_j$ is the $j^{th}$ change point, $\tau_0=0$ and $\tau_{R+1}=m$. Suppose that the first change point in the mean of observations which may be detected is located in the $m_1^{th}$ observation such that $m_1<m$ and $m_1+m_2=m$.

Log of likelihood function for $x_t$ is

$$\log_e \lambda - \lambda x_t, \tag{4}$$

and log of likelihood function for the first $m_1$ observations is

$$\log_e L_1 = m_1 \log_e \lambda_1 - \lambda_1 \sum_{t=1}^{m_1} x_t . \tag{5}$$

This term is maximized when

$$\frac{1}{\overline{x}_1} = \frac{m_1}{\sum_{t=1}^{m_1} x_t} \tag{6}$$

which is the maximum likelihood estimator (MLE) for the first $m_1$ observations. The maximized value of the LRT statistic is then

$$L_1 = m_1 \log_e \left(\frac{m_1}{\sum_{t=1}^{m_1} x_t}\right) - m_1 . \tag{7}$$

Similarly, the likelihood function for the remaining $m_2 = m - m_1$ observations is maximized when

$$\frac{1}{\bar{x}_2} = \frac{m_2}{\sum_{t=m-m_1+1}^{m} x_t} \qquad (8)$$

with a value of

$$L_2 = m_2 \log_e\left(\frac{m_2}{\sum_{t=m-m_1+1}^{m} x_t}\right) - m_2. \qquad (9)$$

Hence, under the alternative hypothesis $H_a$ which state that there is at least a change point in the random input data set, the maximum log-likelihood function for all observations is the sum of the two log-likelihood functions

$$L_a = L_1 + L_2. \qquad (10)$$

Conversely, if all $m$ observations in the input data set are independently and identically distributed then the likelihood function is maximized when

$$\frac{1}{\bar{x}} = \frac{m}{\sum_{t=1}^{m} x_t} \qquad (11)$$

with a value of

$$L_0 = m \log_e\left(\frac{m}{\sum_{t=1}^{m} x_t}\right) - m. \qquad (12)$$

If $L_a$ is considerably larger than $L_0$ we could conclude $H_a$ with $100(1-\alpha)$ percent, i.e. our input data are not homogeneous over the whole time span in which they were being gathered. This enables us to statistically diagnose the presence of nonhomogeneity in random input data set $S=\{X_t; 1,2,\ldots,m\}$ and also provides a base for estimating the exact length of subintervals more accurately. It should be pointed out that

$$\begin{aligned} \text{lrt}[m_1, m_2] &= -2(L_0 - L_a) \\ &= 2\left[m \log_e \bar{x} - m_1 \log_e \bar{x}_1 - m_2 \log_e \bar{x}_2\right] \\ &= 2 \log_e \left[\left(\frac{\bar{x}}{\bar{x}_1}\right)^{m_1} \left(\frac{\bar{x}}{\bar{x}_2}\right)^{m_2}\right] \end{aligned} \qquad (13)$$

has asymptotically a chi square distribution with one degree of freedom, with large values signaling nonhomogeneous input data. For the large sample approximation see Wilks (1947, p. 151) or Mood, Graybill, and Boes (1974, p. 441).

Clearly, the value of $m_1$, the number of observations in the first group, which maximizes equation (13) is the maximum likelihood estimate for the change point location, provided that one exists. Hence, maximum value of equation (13) beyond a predefined boundary indicates that input observations are not all from an identical distribution.

At this point, we shall consider the behavior of the statistic in equation (13) briefly. Based on 4000 simulation runs each with $m$ observations derived from a homogeneous exponential distribution, Elrt $[m_1, m_2]$, the estimated expected value of equation (13) is calculated for each value of $m_1$ using

$$\text{Elrt}[m_1, m_2] = E\left[\text{lrt}[m_1, m_2]\right]. \qquad (14)$$

Without loss of generality, it is assumed that $m$ is 50 and the homogeneous distribution has a mean equal to $1/\lambda$. The results (not reported here) imply that the homogeneous expected value of equation (13) is not the same for all values of $m_1$. If $m_1$ or $m_2$ is small, the expected value is always larger than when they get close to each other. In fact, expected value of likelihood ratio tests is likely to be shaped as a bathtub with heavy tails. Therefore, it is desirable to improve the statistic in (13) by dividing each statistic by its relevant homogeneous expected value, i.e.

$$lrt^* = \frac{\text{lrt}[m_1, m_2]}{\text{Elrt}[m_1, m_2]}. \tag{15}$$

In this way the resulting expected value is the same for all values of $m_1$. So we apply the improved test statistic in (15) instead of equation (13) in our study. In the next section, we statistically compare the efficacy of our proposed methods using Monte Carlo simulation.

## 4. Numerical Comparisons

In this section, we use Mont Carlo simulation to make performance comparisons between the two change point estimators. Researchers often use two different categories of measures for comparing the efficiency of change point detection techniques namely accuracy performance that shows how close a measured value is to the actual value and precision performance that rates how close the measured values are to each other. To completely investigate the performance of the proposed change point estimator, we report measures evaluating both categories for the case that there are multiple changes in random input data $S$. It is assumed that input observations are univariate exponential random variables and there are $R$ changes in the data set. The rate parameter in each group is identical but they differ between consecutive groups. Presume $\Lambda = \{\mu_{j+1} \mid \mu_{j+1} = \mu_j + \delta \times (-1)^j; \mu_j = 1/\lambda_j; j=0,1,\ldots,R-1\}$ is the set of change values where $\lambda_0$ is a predefined initial value and $\delta$ is the magnitude of change. $\Lambda$ is defined such that the difference in rate parameter for two consecutive groups is identical and equal to $\delta$. For example, imagine that there are $R=4$ groups with different rate parameter values and let $\lambda_0 = 1$ and $\delta = 3$. In this case, sequence $\Lambda$ is defined as $\Lambda = \{1, 4, 1, 4\}$. We also consider equally spaced changes alternating between two groups. For example, a single change would be midway in the data sets, and two shifts would be after one-third and two-third of the data sets, etc. One could consider an alternative model in which the change locations would be specified randomly (Turner (2001)).

It is assumed that there are $m=200$ observations from a simulation input variable and it is of interest to detect any changes in random data set $S$ and their locations. The number of changes are set equal to $R=1, 2, 3, 4$ and $\tau_j$; $j=1,\ldots,R$ is the true location of the $j^{th}$ change. Thus, if there is a single change in the data set, then $R=1$, $\tau_1 = 101$ and there are two different groups. The first group consists of the first 100 observations and the second group consists of the second 100 observations. Similarly, if there are four changes in the data set, $R=4$ then $\tau_1$ to $\tau_4$ are namely 40, 80, 120, and 160. In this case, there are five different groups: the first group consisting of the first observation to the $40^{th}$ observation, the second group consisting of the $41^{st}$ observation to the $80^{th}$ observations and so on. To estimate $\tau_j$'s, 1000 replications are used in each simulation run. Both change point estimators can be designed to be capable of detecting at most seven potential changes. However, one could design both change point detection techniques to be capable of detecting and estimating more or less changes. Considering at most seven changes in a data set of size 200, we can evaluate the first seven $d_j^*$'s in clustering method and seven consecutive segments in the LRT method.

For the clustering method, the threshold values for $d_1^*$ to $d_7^*$ are set equal to 0.7686, 0.9435, 0.7571, 0.8119, 0.7343, 0.7369 and 0.6911, respectively leading to probability of false detection values (type I error) namely 0.03, 0.02, 0.02 0.01, 0.01, 0.01, and 0.01. In this case, according to Bonferroni inequality, the overall probability of false detection ($\alpha_{overall}$) cannot exceed 0.11. The threshold values were calculated via Mont Carlo simulation by generating 100 sets each consisting of 100 values of $d_1^*$ to $d_7^*$ and estimating the 100(1- $\alpha$) percentiles of $d_1^*$ to $d_7^*$ where there is no change in the data sets. It is also worth mentioning, we apply the Box-Cox transformation for $x_t$'s as shown in equation (3) to stabilize the standard deviation of input variables and increase the performance of the clustering method. Using 100,000 observations, η was estimated to be 0.24. Therefore, all observations transform into $x^{0.24}$. Also we take into account the effects that may caused by not using the Box-Cox transformation and also by using top-down (divisive) hierarchical clustering algorithm instead of bottom-up method.

Also for the LRT method, we first estimate Elrt array for all possible values of $m_1$ using 4,000 simulation runs while $m=200$ and then we utilize the improved statistic in equation (15). If $lrt_1^*$ exceeds its corresponding threshold value, the data are separated at the location of $lrt_1^*$. Then the method is repeated for two new subsets and is continued for at most $2^0+2^1+2^2=7$ segments. The threshold values for $lrt_1^*$ to $lrt_7^*$ are set equal to 0.7686, 0.9435, 0.7571, 0.8119, 0.7343, 0.7369, and 0.6911, respectively leading to probability of false detection values (type I error) namely 0.03, 0.02, 0.02 0.01, 0.01, 0.01, and 0.01. Similarly, the threshold values were given by estimating the $lrt^*$'s percentiles using simulation.

### 4.1. *Accuracy performances of change point estimators*

Suppose $\hat{\tau}_j^c$ and $\hat{\tau}_j^m$ are the estimates of $\tau_j$ and $j^{th}$ change point derived by clustering and LRT methods, respectively. We compare the accuracy of $\hat{\tau}_j^c$ and $\hat{\tau}_j^m$ when there are $R$ changes in data sets of input observations. Table 1 shows the estimate of change point(s) obtained by clustering method and their standard errors (in the parenthesis) where there are $R=1$, 2, 3, and 4 changes in the simulation input data. As mentioned before, we investigate the seven consecutive $d_j$'s and select the last $j^{th}$ which exceeds its threshold value. If $d_q$; $q \leq 7$, is the last $d_j$'s that is greater than its threshold value, then $l_j$; $j=1,2,…,q$ is considered as the estimate of the $j^{th}$ change point location. Because there are $R$ changes in the input data set, the value of $q > R$ indicates one or more false detections. On the other hand, if $q < R$, the method cannot detect all existence changes that is failing to detect a true change point and identify its location. In Table 1, we consider the $R$ first change point locations $l_j$'s, where $q > R$ and the $q$ first change point locations where $q < R$. It can be observed that the hierarchical clustering method performs properly when the magnitude of change in the mean is greater than 1 ($\delta>1$). For small changes in the mean value, the clustering method has a bias which increases as the magnitude of change decreases. For example, if there are four groups in an input data set (three changes at locations 50, 100, and 150), the estimated change points are namely 59.0, 102.6 and 126.6 when magnitude of shift is 0.5. Similarly, for $\delta=1$, the estimated change locations are 58.7, 98.7 and 139.3 respectively. However, for larger magnitude of changes in rate parameter ($\delta>1$), the estimates of change locations are approximately unbiased. For example, for $R=3$ and $\delta=4$, $\tau_1=50$, $\tau_2=100$ and $\tau_3=150$ the change point locations are given as 51.4, 98.7 and 151.4 respectively which are much closer to the true values. It is worth mentioning that hierarchical clustering method tends to overestimate the locations of the first changes and underestimate the locations of the last changes. For example, if there is four changes in the input data set, $\tau_1=40$ and $\tau_2=80$ are overestimated for all values of δ while $\tau_3=120$ and $\tau_4=160$ are underestimated. For the case that no transformation is employed in the data, we observe that (not shown here) as change magnitude decreases, the correct location of change estimates as well as their standard error are adversely affected. For example, for $R=3$ and δ=1, where $\tau_1=50$, $\tau_2=100$ and $\tau_3=150$, the estimates are 52.8, 96.3, and 154.8 with standard deviations 0.23, 0.21, and 0.19 respectively. Also applying divisive (top-down) hierarchical clustering to the same data, we realize a very small change in most of change point estimates but not their standard errors compared to bottom-up algorithm.

[Insert Table 1 about here]

[Insert Table 2 about here]

Table 2 displays the estimates of change locations based on the LRT method with their related standard error (in the parenthesis) when there are $R+1$ subgroups differing in rate parameter. Note that the LRT method performs effectively when there is a single change point in the input data set and provides unbiased estimates for change location. This method performs appropriately as well if there are more than one changes and the magnitude of shift is larger than one. Although there is a small biasness, particularly in intermediate change points, the magnitude of biasness is not very large to seriously affect groups. Comparison of Table 1 and Table 2 indicates that LRT estimates are superior to clustering method in terms of average and standard error of change points.

As an illustration, assuming $R=4$ and δ=0.5, the estimates of $\tau_1=50$, $\tau_2=100$ and $\tau_3=150$ using clustering method are 59.4, 102.6 and 126.6, respectively with corresponding standard error values of 0.73, 1.52 and 1.31 while LRT estimates are 55.4, 99.9 and 150.8, respectively with corresponding standard error values of 0.25, 0.36 and 0.32. In spite of superiority of LRT method, it should be noted that this method requires the knowledge about the exact distribution of input data set. This assumption makes the method somehow restricted and degrades its practicability. The exact distribution of an input data set cannot generally be determined without any preliminary data analysis. Besides, in most cases, the simulation input data set may follow different distributions in which the observations between groups follow distributions with various functional forms. Thus, the LRT method should be developed from a more generalized form that is flexible for such changes. For example, one can derive the LRT statistic of an exponential family of distributions that can conveniently fit with separate data sets. For other input densities like Johnson or Bezier systems of distributions, in which LRT extension are not straightforward, alternative non-parametric approaches should instead be considered.

### 4.2. *Precision performances of change point estimators*

Researchers often use precision performance measures along with accuracy performance measures and believe the average of change points cannot solely provide a comprehensive comparison of different estimators. An estimator with good performances in estimating location of changes may inherently have poor performances in terms of dispersion. In this situation, the estimator provides estimates that are close to the true location in average but far from each other. An indication of the precision of two estimators is confidence set that is the observed frequency with which the change point estimators were within a given number of subgroups of the actual change point.

[Insert Table 3 about here]

Table 3 gives the estimated precision performances over a range of δ for $\hat{\tau}_j^c$, the estimate of $\tau_j$; $j=1,…,R$ obtained by the clustering method. It is perceived that the precision performance of the hierarchical clustering method is poor particularly when δ is small. For instance, only 4% of all estimates of $\tau_1$ show the true change value when there is a change of size 1 in the mean value and 13% of all estimates are 2 units or less far from the true change. Furthermore, 59% of all estimates are 25 units or more from the true change when δ=1 indicating a grave problem in precision performance of the clustering method when the magnitude of change is small. As it can be seen from Table 3, the precision performance of the estimates, obtained by the clustering method raises as the magnitude of change increase. For example, in 90% of time, a change of size δ ≥ 4 is detected at least in a location five or less units far from the true change point. Therefore, the method can appropriately be used when the change of interest is large or moderate. For small magnitude of changes other estimators e.g. the LRT estimates are suggested. In case of not using Box-Cox transformation, the precision measure is adversely influenced especially with small magnitudes, and gradually improves as magnitude of change increases. However, similar to accuracy performance metric, changing agglomerative algorithm to divisive has very little impact on the results.

[Insert Table 4 about here]

Table 4 presents the precision performance values given by the LRT method over different values of δ. It is shown that although a change of size δ=0.5 may be far from the true location of change, the method totally have an appropriate performance. For example, a small change of size δ=0.5 can be detected about five or less units from the true change point $\tau_1$. Besides, the method can guarantee to identify a location equal to or less than 15 units from the true location approximately in 90% of times. When compared to the hierarchical clustering method, the LRT estimates of change points perform seemly and provide reliable inferences for the input simulation data. We should point out that although both methods are often unable to perform perfectly but both methods often provide stable groups that are identical. This is expected to occur due to the fact that they typically divide data set into groups which include a large number of identical observations following a major distribution and relatively few observations following a minor distribution. These minor observations are not often large enough to reject the stability hypotheses and cannot seriously hurt the output results.

## 5. A Case Example

In this section we consider the well-known Able Baker Carhop problem which is simply an M/G/2 queueing system with heterogeneous servers (Banks et al., 2000) to show the application of the proposed method. Imagine a drive-in restaurant with two service channel where carhops take orders and bring food to the car. As depicted in Figure 2, there are two carhops–Able and Baker. Cars come in the manner that the interarrival times can be modeled as an exponential random variable. Able is more skillful to do the job and hence works a little faster than Baker. The distribution of their service times is suitably estimated from historical data; for Able it is a triangular random variable with a range from 5 to 10 minutes and a mean of 7 minutes and for Baker it is also a triangular random variable but with a range from 6 to 11 minutes and a mean of 8 minutes. It is assumed that Able has seniority so can get the customer if both carhops are idle. In other words, the logic of the system when a customer arrives is as follows: If the customer finds Able idle, his/her service is started right away by Able. If Able is busy and Baker is idle, the customer begins service immediately with Baker. If both are not idle, the customer starts service with the first server to become free. Note that the restaurant is empty at the beginning of each simulation run and after service completion the customer leaves the system.

[Insert Figure 2 about here]

Caption: Figure 2. Able & Baker Carhop Model

To effectively serve our purpose, we consider two different modes of arrival times: (I) the interarrival times of the first 50 cars come from an exponential random variable with mean of 1 minute. Then starting from the 51[th] car up to the 100[th], they are created from the same random variate but with mean of 6 ($\mu_{j+1} = \mu_j + \delta \times (-1)^j$; j=2). Hence, $\mu_3$ and $\mu_4$, the mean of third and fourth group of 50 cars, are 1 and 6 minutes, respectively. (II) All of 200 interarrival times generated in case (I) are grouped as one batch with pooled mean of 3.329. Note that in order to obtain the best-fit model in case (II), we perform Anderson-Darling goodness of fit test (Anderson and Darling (1952)). One may use different methods discussed in the literature to fit various models to this nonhomogeneous batch.

To compare the system measures of performance, 100 simulations were conducted and the brief results are as follows:

In *case (I)*:

1. Both Able and Baker were busy 54% of the times on average.

2. The mean number of entities in the system was 17.69 with a maximum of 72 customers for replication number 9.

3. Able can serve around 106.51 customers while Baker can seize 93.49 customers averagely on each simulation run.

In *case (II)*:

1. Able and Baker were on the average busy nearly 65.4% and 64.8% of the times respectively.

2. The mean number of entities in the system was 9.43 with a maximum of 34 customers for replication number 40.

3. Able can serve around 106.6 customers while Baker can seize 93.4 customers averagely.

Although these two models are similar in the mean number of customers seized by each server, they differ not only in long-time average number of customers present in the system but also in utilization percent of resources. Hence, in order to avoid misleading interpretation of the outputs of systems studied, one has to first check the homogeneity of collected input data.

## 6. Conclusion

In this paper, two methods were proposed for the case that the simulation input data is time-dependent Poisson process but can be clustered in identical groups. The methods deal with identifying the locations (groups) in such a way that the observations within each group follow the same probability distribution but observations in two consecutive groups have different distributions. Our most important contributions in this work can be summarized as first using a new distance function for the hierarchical clustering method with a sound property under parametric conditions and second presenting a modification format of LRT technique accounted for skewed densities. It was shown that the first method, the clustering technique, can provide estimates without considering any distributional assumptions, however, the estimates does not have appropriate performances. On the other hand, the second method based on the likelihood ratio test has superior accuracy and precision performances while the method still relies on relatively tight distributional assumptions. Modifications of such obstacles can be considered as a good future research. For instance, more efficient distance measures can be used in the hierarchical clustering method leading to improved performances. Furthermore, the LRT method can be generalized by obtaining the LRT statistic for more general densities such as exponential family distributions.

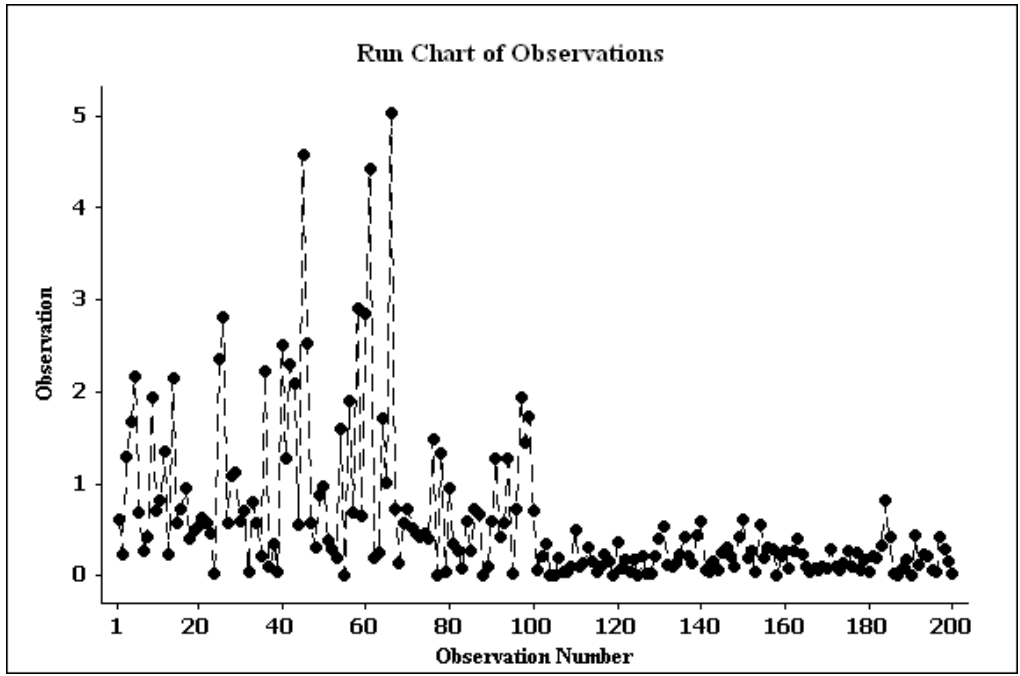

Figure 1. Run Chart of Observations

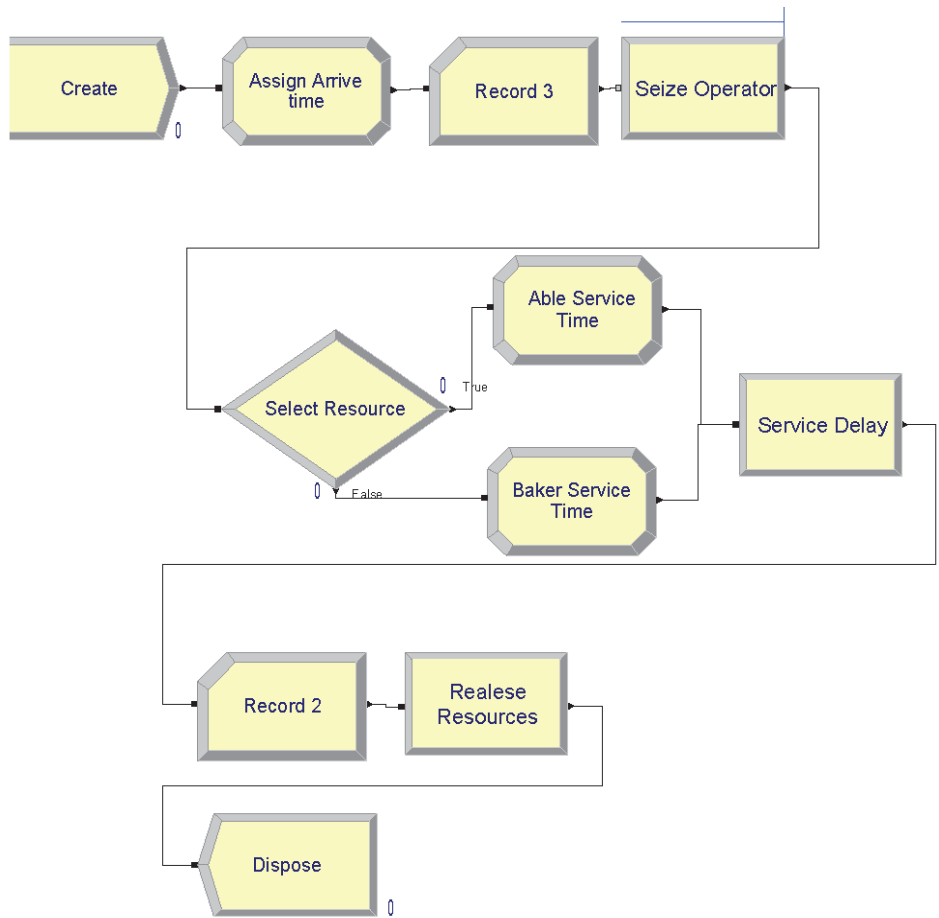

Figure 2. Able & Baker Carhop Model

Table 1: Accuracy performances for the clustering method when there are $R$ changes in locations $\tau_j$; $j=1,\ldots,R$

| | Number of change points | | | | | | | | | |
|---|---|---|---|---|---|---|---|---|---|---|
| | $R=1$ | $R=2$ | | $R=3$ | | | $R=4$ | | | |
| | $\tau_1=100$ | $\tau_1=66$ | $\tau_2=133$ | $\tau_1=50$ | $\tau_2=100$ | $\tau_3=150$ | $\tau_1=40$ | $\tau_2=80$ | $\tau_3=120$ | $\tau_3=160$ |
| $\delta$ | $\hat{\tau}_1^c$ | $\hat{\tau}_1^c$ | $\hat{\tau}_2^c$ | $\hat{\tau}_1^c$ | $\hat{\tau}_2^c$ | $\hat{\tau}_3^c$ | $\hat{\tau}_1^c$ | $\hat{\tau}_2^c$ | $\hat{\tau}_3^c$ | $\hat{\tau}_4^c$ |
| 0.5 | 109.5 (1.20) | 76.5 (0.94) | 104.3 (1.29) | 59.0 (0.73) | 102.6 (1.52) | 126.6 (1.31) | 47.3 (0.57) | 86.6 (1.24) | 110.0 (1.13) | 134.7 (1.27) |
| 1 | 107.3 (0.72) | 77.1 (0.71) | 116.2 (0.88) | 58.5 (0.55) | 98.7 (1.07) | 139.3 (0.97) | 46.7 (0.47) | 84.1 (1.09) | 115.0 (0.93) | 153.1 (0.83) |
| 2 | 102.5 (0.25) | 69.1 (0.28) | 129.6 (0.30) | 54.0 (0.32) | 96.8 (0.41) | 150.7 (0.48) | 44.0 (0.29) | 77.2 (0.46) | 121.5 (0.46) | 157.4 (0.47) |
| 3 | 101.7 (0.14) | 67.8 (0.14) | 131.4 (0.15) | 51.9 (0.16) | 97.8 (0.18) | 152.1 (0.18) | 42.0 (0.17) | 78.1 (0.18) | 121.8 (0.19) | 157.7 (0.19) |
| 4 | 101.3 (0.11) | 67.2 (0.10) | 131.8 (0.11) | 51.4 (0.11) | 98.7 (0.12) | 151.4 (0.11) | 41.5 (0.12) | 78.4 (0.12) | 121.3 (0.12) | 158.4 (0.12) |
| 5 | 101.1 (0.09) | 67.0 (0.08) | 131.8 (0.09) | 51.1 (0.09) | 98.7 (0.09) | 151.0 (0.09) | 41.2 (0.09) | 78.8 (0.09) | 121.2 (0.09) | 158.8 (0.10) |

Table 2: Accuracy performances for the LRT method when there are R changes in locations τj; $j=1,\ldots,R$

| | Number of change points | | | | | | | | | |
|---|---|---|---|---|---|---|---|---|---|---|
| | $R=1$ | $R=2$ | | $R=3$ | | | $R=4$ | | | |
| | $\tau_1=100$ | $\tau_1=66$ | $\tau_2=133$ | $\tau_1=50$ | $\tau_2=100$ | $\tau_3=150$ | $\tau_1=40$ | $\tau_2=80$ | $\tau_3=120$ | $\tau_3=160$ |
| $\delta$ | $\hat{\tau}_1^c$ | $\hat{\tau}_1^c$ | $\hat{\tau}_2^c$ | $\hat{\tau}_1^c$ | $\hat{\tau}_2^c$ | $\hat{\tau}_3^c$ | $\hat{\tau}_1^c$ | $\hat{\tau}_2^c$ | $\hat{\tau}_3^c$ | $\hat{\tau}_4^c$ |
| 0.5 | 100.7 (0.34) | 71.9 (0.30) | 132.0 (0.43) | 55.4 (0.25) | 99.9 (0.36) | 150.8 (0.32) | 31.2 (0.52) | 78.2 (0.32) | 123.4 (0.28) | 161.4 (0.31) |
| 1 | 101.2 (0.25) | 69.2 (0.25) | 133.9 (0.30) | 53.0 (0.19) | 98.8 (0.29) | 153.0 (0.24) | 36.7 (0.35) | 78.3 (0.23) | 123.2 (0.24) | 160.6 (0.23) |
| 2 | 100.9 (0.15) | 66.9 (0.15) | 133.6 (0.18) | 51.1 (0.12) | 97.7 (0.18) | 152.2 (0.18) | 39.4 (0.20) | 76.6 (0.19) | 122.6 (0.17) | 160.9 (0.15) |
| 3 | 100.6 (0.09) | 66.4 (0.11) | 133.6 (0.14) | 50.8 (0.12) | 97.7 (0.16) | 153.0 (0.16) | 39.8 (0.14) | 76.5 (0.18) | 123.1 (0.16) | 160.8 (0.12) |
| 4 | 100.7 (0.09) | 66.5 (0.10) | 133.5 (0.14) | 50.4 (0.09) | 97.5 (0.14) | 152.9 (0.16) | 40.0 (0.10) | 76.2 (0.17) | 123.7 (0.16) | 160.4 (0.09) |
| 5 | 100.4 (0.07) | 66.4 (0.10) | 133.2 (0.10) | 50.5 (0.10) | 97.5 (0.14) | 152.6 (0.15) | 40.2 (0.11) | 75.8 (0.16) | 123.8 (0.15) | 160.3 (0.08) |

Table 3: Estimated precision performances for the clustering method when there are $R$ changes in locations $\tau_j$; $j=1,\ldots,R$

| | | Clustering Method | | | | | | |
|---|---|---|---|---|---|---|---|---|
| | $\delta$ | $\hat{P}(|\hat{\tau}_1^c - \tau_1|)=0$ | $\hat{P}(|\hat{\tau}_1^c - \tau_1|)\leq 1$ | $\hat{P}(|\hat{\tau}_1^c - \tau_1|)\leq 2$ | $\hat{P}(|\hat{\tau}_1^c - \tau_1|)\leq 5$ | $\hat{P}(|\hat{\tau}_1^c - \tau_1|)\leq 10$ | $\hat{P}(|\hat{\tau}_1^c - \tau_1|)\leq 15$ | $\hat{P}(|\hat{\tau}_1^c - \tau_1|)\leq 25$ |
| $\tau_1$ | 0.5 | 0.00 | 0.02 | 0.02 | 0.05 | 0.07 | 0.10 | 0.13 |
| | 1 | 0.04 | 0.09 | 0.13 | 0.21 | 0.28 | 0.35 | 0.41 |
| | 2 | 0.18 | 0.32 | 0.43 | 0.61 | 0.75 | 0.83 | 0.91 |
| | 3 | 0.27 | 0.47 | 0.60 | 0.80 | 0.92 | 0.96 | 0.99 |
| | 4 | 0.39 | 0.58 | 0.70 | 0.88 | 0.97 | 0.99 | 1.00 |
| | 5 | 0.43 | 0.65 | 0.77 | 0.93 | 0.98 | 1.00 | 1.00 |
| | | $\hat{P}(|\hat{\tau}_2^c - \tau_2|)=0$ | $\hat{P}(|\hat{\tau}_2^c - \tau_2|)\leq 1$ | $\hat{P}(|\hat{\tau}_2^c - \tau_2|)\leq 2$ | $\hat{P}(|\hat{\tau}_2^c - \tau_2|)\leq 5$ | $\hat{P}(|\hat{\tau}_2^c - \tau_2|)\leq 10$ | $\hat{P}(|\hat{\tau}_2^c - \tau_2|)\leq 15$ | $\hat{P}(|\hat{\tau}_2^c - \tau_2|)\leq 25$ |
| $\tau_2$ | 0.5 | 0.01 | 0.02 | 0.03 | 0.04 | 0.06 | 0.08 | 0.12 |
| | 1 | 0.05 | 0.10 | 0.12 | 0.19 | 0.26 | 0.31 | 0.39 |
| | 2 | 0.19 | 0.33 | 0.42 | 0.60 | 0.75 | 0.84 | 0.90 |
| | 3 | 0.34 | 0.52 | 0.65 | 0.82 | 0.92 | 0.97 | 0.99 |
| | 4 | 0.36 | 0.55 | 0.66 | 0.86 | 0.97 | 0.99 | 1.00 |
| | 5 | 0.45 | 0.64 | 0.75 | 0.91 | 0.99 | 1.00 | 1.00 |
| | | $\hat{P}(|\hat{\tau}_3^c - \tau_3|)=0$ | $\hat{P}(|\hat{\tau}_3^c - \tau_3|)\leq 1$ | $\hat{P}(|\hat{\tau}_3^c - \tau_3|)\leq 2$ | $\hat{P}(|\hat{\tau}_3^c - \tau_3|)\leq 5$ | $\hat{P}(|\hat{\tau}_3^c - \tau_3|)\leq 10$ | $\hat{P}(|\hat{\tau}_3^c - \tau_3|)\leq 15$ | $\hat{P}(|\hat{\tau}_3^c - \tau_3|)\leq 25$ |
| $\tau_3$ | 0.5 | 0.01 | 0.01 | 0.01 | 0.03 | 0.05 | 0.05 | 0.07 |
| | 1 | 0.04 | 0.08 | 0.11 | 0.18 | 0.24 | 0.28 | 0.34 |
| | 2 | 0.18 | 0.34 | 0.45 | 0.63 | 0.76 | 0.82 | 0.89 |
| | 3 | 0.33 | 0.52 | 0.65 | 0.82 | 0.92 | 0.96 | 0.99 |
| | 4 | 0.37 | 0.59 | 0.70 | 0.88 | 0.97 | 0.99 | 1.00 |
| | 5 | 0.42 | 0.64 | 0.75 | 0.93 | 0.99 | 1.00 | 1.00 |
| | | $\hat{P}(|\hat{\tau}_4^c - \tau_4|)=0$ | $\hat{P}(|\hat{\tau}_4^c - \tau_4|)\leq 1$ | $\hat{P}(|\hat{\tau}_4^c - \tau_4|)\leq 2$ | $\hat{P}(|\hat{\tau}_4^c - \tau_4|)\leq 5$ | $\hat{P}(|\hat{\tau}_4^c - \tau_4|)\leq 10$ | $\hat{P}(|\hat{\tau}_4^c - \tau_4|)\leq 15$ | $\hat{P}(|\hat{\tau}_4^c - \tau_4|)\leq 25$ |
| $\tau_4$ | 0.5 | 0.00 | 0.00 | 0.00 | 0.01 | 0.01 | 0.01 | 0.02 |
| | 1 | 0.02 | 0.06 | 0.08 | 0.13 | 0.17 | 0.19 | 0.23 |
| | 2 | 0.17 | 0.32 | 0.43 | 0.61 | 0.75 | 0.81 | 0.87 |
| | 3 | 0.29 | 0.48 | 0.61 | 0.81 | 0.91 | 0.96 | 0.99 |
| | 4 | 0.36 | 0.57 | 0.72 | 0.88 | 0.97 | 0.99 | 1.00 |
| | 5 | 0.44 | 0.64 | 0.76 | 0.91 | 0.98 | 1.00 | 1.00 |

Table 4: Estimated precision performances for the LRT method when there are R changes in locations $\tau_j$; $j=1,\ldots,R$

| | | LRT Method | | | | | | |
|---|---|---|---|---|---|---|---|---|
| | $\delta$ | $\hat{P}(|\hat{\tau}_1^c - \tau_1|)=0$ | $\hat{P}(|\hat{\tau}_1^c - \tau_1|)\leq 1$ | $\hat{P}(|\hat{\tau}_1^c - \tau_1|)\leq 2$ | $\hat{P}(|\hat{\tau}_1^c - \tau_1|)\leq 5$ | $\hat{P}(|\hat{\tau}_1^c - \tau_1|)\leq 10$ | $\hat{P}(|\hat{\tau}_1^c - \tau_1|)\leq 15$ | $\hat{P}(|\hat{\tau}_1^c - \tau_1|)\leq 25$ |
| $\tau_1$ | 0.5 | 0.10 | 0.14 | 0.17 | 0.31 | 0.43 | 0.57 | 0.79 |
| | 1 | 0.15 | 0.32 | 0.44 | 0.64 | 0.76 | 0.83 | 0.94 |
| | 2 | 0.26 | 0.49 | 0.63 | 0.80 | 0.90 | 0.95 | 0.99 |
| | 3 | 0.37 | 0.62 | 0.76 | 0.88 | 0.95 | 0.98 | 1.00 |
| | 4 | 0.46 | 0.73 | 0.84 | 0.93 | 0.97 | 0.99 | 1.00 |
| | 5 | 0.49 | 0.72 | 0.83 | 0.94 | 0.96 | 0.98 | 1.00 |
| | | $\hat{P}(|\hat{\tau}_2^c - \tau_2|)=0$ | $\hat{P}(|\hat{\tau}_2^c - \tau_2|)\leq 1$ | $\hat{P}(|\hat{\tau}_2^c - \tau_2|)\leq 2$ | $\hat{P}(|\hat{\tau}_2^c - \tau_2|)\leq 5$ | $\hat{P}(|\hat{\tau}_2^c - \tau_2|)\leq 10$ | $\hat{P}(|\hat{\tau}_2^c - \tau_2|)\leq 15$ | $\hat{P}(|\hat{\tau}_2^c - \tau_2|)\leq 25$ |
| $\tau_2$ | 0.5 | 0.04 | 0.13 | 0.18 | 0.33 | 0.62 | 0.73 | 1.00 |
| | 1 | 0.16 | 0.27 | 0.37 | 0.62 | 0.80 | 0.92 | 1.00 |
| | 2 | 0.22 | 0.43 | 0.54 | 0.70 | 0.84 | 0.94 | 1.00 |
| | 3 | 0.25 | 0.42 | 0.54 | 0.72 | 0.88 | 0.95 | 1.00 |
| | 4 | 0.31 | 0.46 | 0.59 | 0.73 | 0.87 | 0.95 | 1.00 |
| | 5 | 0.25 | 0.42 | 0.53 | 0.70 | 0.89 | 0.96 | 1.00 |
| | | $\hat{P}(|\hat{\tau}_3^c - \tau_3|)=0$ | $\hat{P}(|\hat{\tau}_3^c - \tau_3|)\leq 1$ | $\hat{P}(|\hat{\tau}_3^c - \tau_3|)\leq 2$ | $\hat{P}(|\hat{\tau}_3^c - \tau_3|)\leq 5$ | $\hat{P}(|\hat{\tau}_3^c - \tau_3|)\leq 10$ | $\hat{P}(|\hat{\tau}_3^c - \tau_3|)\leq 15$ | $\hat{P}(|\hat{\tau}_3^c - \tau_3|)\leq 25$ |
| $\tau_3$ | 0.5 | 0.04 | 0.08 | 0.27 | 0.40 | 0.69 | 0.90 | 1.00 |
| | 1 | 0.12 | 0.26 | 0.33 | 0.56 | 0.77 | 0.91 | 1.00 |
| | 2 | 0.22 | 0.41 | 0.53 | 0.72 | 0.91 | 0.98 | 1.00 |
| | 3 | 0.26 | 0.44 | 0.56 | 0.74 | 0.90 | 0.96 | 1.00 |
| | 4 | 0.28 | 0.45 | 0.54 | 0.72 | 0.87 | 0.97 | 1.00 |
| | 5 | 0.26 | 0.41 | 0.53 | 0.72 | 0.88 | 0.96 | 1.00 |
| | | $\hat{P}(|\hat{\tau}_4^c - \tau_4|)=0$ | $\hat{P}(|\hat{\tau}_4^c - \tau_4|)\leq 1$ | $\hat{P}(|\hat{\tau}_4^c - \tau_4|)\leq 2$ | $\hat{P}(|\hat{\tau}_4^c - \tau_4|)\leq 5$ | $\hat{P}(|\hat{\tau}_4^c - \tau_4|)\leq 10$ | $\hat{P}(|\hat{\tau}_4^c - \tau_4|)\leq 15$ | $\hat{P}(|\hat{\tau}_4^c - \tau_4|)\leq 25$ |
| $\tau_4$ | 0.5 | 0.04 | 0.10 | 0.21 | 0.46 | 0.71 | 0.85 | 1.00 |
| | 1 | 0.13 | 0.32 | 0.44 | 0.63 | 0.85 | 0.93 | 1.00 |
| | 2 | 0.24 | 0.50 | 0.61 | 0.80 | 0.94 | 0.99 | 1.00 |
| | 3 | 0.34 | 0.63 | 0.76 | 0.90 | 0.97 | 0.99 | 1.00 |
| | 4 | 0.44 | 0.69 | 0.80 | 0.94 | 0.99 | 1.00 | 1.00 |
| | 5 | 0.48 | 0.76 | 0.86 | 0.95 | 0.99 | 1.00 | 1.00 |